\documentstyle[12pt]{article}
\begin{document}

\def \inbar{\vrule height1.5ex width.4pt depth0pt}
\def \C{\relax\hbox{\kern.25em$\inbar\kern-.3em{\rm C}$}}
\def \R{\relax{\rm I\kern-.18em R}}
\newcommand{\Z}{\ Z \hspace{-.08in}Z}
\newcommand{\be}{\begin{equation}}
\newcommand{\ee}{\end{equation}}
\newcommand{\bea}{\begin{eqnarray}}
\newcommand{\eea}{\end{eqnarray}}
\newcommand{\nn}{\nonumber}
\renewcommand{\ll}{\left[ }
\newcommand{\rr}{\right] }
\newcommand{\kt}{\rangle}
\newcommand{\br}{\langle}
\newcommand{\lll}{\left( }
\newcommand{\rrr}{\right)}
\newcommand{\dagg}{\dagger}
\newcommand{\rdel}{\stackrel{\rightarrow}{\partial}}
\newcommand{\ldel}{\stackrel{\leftarrow}{\partial}}
\newcommand{\pbra}{[\hspace{-.3mm}[}
\newcommand{\pket}{]\hspace{-.3mm}]}

\baselineskip=24pt

\title{Para-Generalization of Peierls Bracket Quantization}
\author{Ali Mostafazadeh\\ \\
Institute for Studies in Theoretical Physics and Mathematics\\
P.~O.~Box 19395-5746, Tehran, Iran, and \\ Department of Physics,
Sharif University of Technology\\ P.~O.~Box 11365-9161, Tehran, Iran.
\thanks{E-mail: ``alim@netware2.ipm.ac.ir,'' Fax: (98-21)228-0415.}}
\maketitle

\vspace{1cm}

\begin{abstract}
A convenient formalism is developed to treat classical dynamical systems
involving $(p=2)$ parafermionic and parabosonic dynamical variables.
This is achieved via the introduction of a parabracket which summarizes
the paracommutation relations of the corresponding Green components
in a unified manner. Furthermore, it is shown that Peierls quantization
scheme may be applied to such systems provided that one uses the
above mentioned parabracket to express the quantum paracommutation
relations. Application of the Peierls scheme also provides the
form of the parafermionic and parabosonic kinetic terms in the Lagrangian.
\end{abstract}

\newpage
\section{Introduction}
Recently, {\em Parasupersymmetry} \cite{r-s,b-d,p8} and {\em Fractional
Supersymmetry} \cite{frac} have been attracting much attention.
This may be best explained by noting the achievements of workers
in {\em Supersymmetry} \cite{susy}. Afterall parasupersymmetry and fractional
supersymmetry may be viewed as generalizations of the ordinary
supersymmetry. This is most easily seen in the structure of the
defining algebraic expressions.

For the case of ($p=2$)-parasupersymmetry, it is shown, in the
most general setting, that the degeneracy structure is almost fully
determined by the defining parasuperalgebra \cite{p8}. In fact, for
a large class of ($p=2$)-parasupersymmetric quantum systems one
can even define the analog of the Witten index \cite{witten}
of supersymmetry \cite{p8}. Like the Witten index, this integer
is a topological invariant linked to the indices of Fredholm (resp.~
elliptic) operators for the known cases \cite{p9}. Physically,
it signifies the exactness or breaking of parasupersymmetry \cite{p8}.

These indications of the similarity between supersymmetry
and parasupersymmetry urges one to seek for a better understanding of both
the classical and quantum versions of parasupersymmtry.
Parasupersymmetric quantum mechanics (PSQM) has been studied to some extent
in the framework of specific quantum mechanical examples
\cite{r-s,b-d,khare}. Its classical counterpart, however, has not
been studied properly, to the best of author's knowledge. A discouraging
factor in such a study would be the complicated algebraic structure of
the associated {\em para-Grassmann variables}. The latter were introduced
in the study of {\em parastatistics} \cite{o-k} which is directly
related with parasupersymmetry.

In 1953, Green \cite{green} proposed a generalization of quantum field
theory that allowed for dynamical fields with generalized statistics
or parastatistics. Such theories were studied in a series of articles
in 60's and 70's before the advent of supersymmetry \cite{g-m,o-k}.
Parastatistics of Green has found some application in string theory
\cite{a-m} and provided an alternative point of view for theories with
internal symmetries \cite{gro}. A thorough reveiw of the subject is
provided in Ref.~\cite{o-k}.

To relate the new ($p=2$)-PSQM with the old parastatistics of Green, one may
begin with a study of its classical analog. The corresponding classical
parasupersymmetric systems will involve para-Grassmann variables $\psi$ of
order 2, i.e., $\psi^3=0$. As is the case for ordinary fermionic variables
($\psi^2=0$), the Lagrangian formulation is most convenient to study such
systems. This observation stems from the fact that fermionic coordinates,
due to the form of their kinetic term in the Lagrangian, are proportional
with their corresponding conjugate momenta. Thus these systems are indeed
constrained and the proper treatment of the constraints is necessary in
their (Hamiltonian)  canonical quantization \cite{goodman}.
The Lagrangian formulation lacks the apparent difficulties with these first
class constraints. A quantization scheme applicable in the framework of
Lagrangian mechanics was proposed by Peierls \cite{peierls} for bosonic
systems, and generalized for fermionic and superclassical systems by
De~Witt \cite{bd}. For a demonstration of the application of this method
to supersymmetric systems see Refs.~\cite{bd,p5}.

The aim of the present article is to provide a simple formalism
which would allow for a concise and unified treatment of both parafermionic
and parabosonic dynamical variables of order 2. This allows for a
development of the Lagrangian formulation of para-classical mechanics and
a generalization of Peierls quantization scheme.
In section~2 a brief reveiw of the algebra of creation and annihilation
operators for parafermionic and parabosonic degrees of freedom
and their classical counterparts is provided.
Section~3 specializes to the case $p=2$. Here
a parabracket is introduced which summarizes the algebra of Green's
components and unifies the treatment of both types of degrees of freedom.
Section~4 first discusses the Peierls bracket quantization scheme for
classical systems involving ordinary fermionic and bosonic variables.
A generaliztion of this approach for ($p=2$) Green's components is then
proposed.  In section~5, the Peierls bracket quantization is applied
to a simple one-dimensional parafermi system. The requirement
of the consistency of the canonical and Peierls quantization methods leads
to the determination of the kinetic term in the Lagrangian. Section~6
includes author's final remarks.

\section{Green's Parastatistics}
As defined in Ref.~\cite{gro}, parafermionic (parabosonic) statistics
of order $p$, is a type of statistics -- called {\em parastatistics} -- which
restricts the number of identical particles in a totally symmetric
(resp.~ antisymmetric) state to be at most $p$. Clearly, for $p=1$,
one recovers the ordinary fermionic or Fermi-Dirac (bosonic or Bose-Einstein)
statistics.

Parastatistics is generally signified with the following set of algebraic
relations \cite{green,g-m,o-k}:
        \bea
        [ a_k,[ a_l^\dagger,a_m]_\mp ] &=&2\delta_{kl}a_m
        \;,\nn\\
        {[} a_k , {[} a_l^\dagger,a_m^\dagger {]_\mp} {]} &=&2\delta_{kl}
        a_m^\dagger\mp 2\delta_{km}a_l^\dagger\;,
\label{q1}\\
        {[} a_k,{[} a_l,a_m{]_\mp}{]} &=&0\;,\nn
        \eea
where $a_k^\dagger$ and $a_k$ denote the creation and annihilation
operators, $\ll x,y\rr_\mp :=xy\mp yx\;,~~\forall x,y$, and the signs
$-$ and $+$ correspond to parafermions and parabosons, respectively.

In general, the order $p$ of parastatistics appears as the label
of a representation of the algebra {\large\bf$a$}, generated by
$a_k$ and $a_k^\dagger$ with the rules (\ref{q1}). An irreducible
representation is provided by choosing a unique vacuum state vector $|0\kt$:
        \[ a_k|0\kt=0\;,~~~~\forall k\;,\]
and constructing a Hilbert-Fock space ${\cal A}$ using the basic vectors:
        \[ |k_1,\cdots,k_l\kt:=a_{k_1}^\dagger\cdots a_{k_l}^\dagger
        |0\kt\;.\]
One can show \cite{g-m,o-k} that in this representation
        \[ a_ka_l^\dagger=p\delta_{kl}|0\kt\;,\]
for some non-negative integer $p$.

In his original article, Green \cite{green} proposed another
(reducible \cite{g-m}) representation of the algebra {\large\bf$a$},
which involved bilinear algebraic relations rather than the complicated
trilinear relations (\ref{q1}). Green defined the algebra
{\large\bf$b$}, generated by the generators $\zeta^\alpha_k$,
and $\zeta^{\alpha\dagger}_k$, $\alpha=0,\cdots,p-1$, and rules:
        \bea
        \ll \zeta_k^\alpha,\zeta^{\alpha\dagger}_j\rr_\mp
        &=&\delta_{kj}\;,\nn\\
        \ll \zeta_k^\alpha,\zeta^{\alpha}_j\rr_\mp
        &=&0\;,
\label{q2}\\
        \ll \zeta_k^\alpha,\zeta^{\beta\dagger}_j\rr_\pm
        &=& \ll \zeta_k^\alpha,\zeta^{\beta}_j\rr_\pm
        \:=\: 0\;,~~~~(\alpha\neq\beta)\,.\nn
        \eea
These relations together with the identification:
        \be
        a_k=\sum_{\alpha=0}^{p-1}\zeta_k^\alpha\;,
\label{q3}
        \ee
lead to the defining relation of {\large\bf$a$}, i.e., Eqs.~(\ref{q1}).

Choosing the same vacuum state vector, $|0\kt$, requiring:
        \[
        \zeta_k^\alpha|0\kt=0\;~~~~\forall \alpha,k\;,\]
and defining a Hilbert-Fock space ${\cal B}$ using the basic vectors:
        \[
        |k_1,\alpha_1;\cdots ;k_m,\alpha_m\kt:=\zeta_{k_1}^{\alpha\dagger}
        \cdots \zeta_{k_m}^{\alpha_m\dagger}|0\kt\;,\]
one obtains a representation of {\large\bf$b$}. In view of Eq.~(\ref{q3}),
this also provides a representation for {\large\bf$a$}. This representation
is known as the {\em Green representation} and $\zeta^\alpha_k$ are
called the {\em Green components} of $a_k$. Note that by construction
the spaces ${\cal A}$ and ${\cal B}$ are in one-to-one correspondence
with the polynomial rings generated by $a_k^\dagger$ and
$\zeta^{\alpha\dagger}_k$, respectively. Thus according to Eq.~(\ref{q3})
${\cal A}$ may be viewed as a subring of ${\cal B}$. In fact, the physical
space to be considered is ${\cal A}$ and not ${\cal B}$. The latter is
introduced for practical convenience.

Another important point is the possibility of the existence of particles
with different types of parastatistics. This is especially the case
for parasupersymmetric systems. The treatment of this case leads to
the introduction of relative parastatistics \cite{g-m}.

Consider two species of particles $a$ and $b$, with creation and
annihilation operators $a_i^\dagger,~a_i$ and $b_j^\dagger,~b_j$,
and orders of parastatistics $p_a$ and $p_b$, respectively. Then,
it can be shown \cite{g-m} that if $p_a\neq p_b$, the
operators $a_i^\dagger$ and $a_i$ either commute or anticommute with
$b_j^\dagger$ and $b_j$. If $p_a=p_b=:p$ then there is a set of
trilinear relations between these operators. The latter can be more
easily expressed in terms of the corresponding Green's components:
        \be
        a_i=\sum_{\alpha=0}^{p-1}\zeta_i^\alpha~,~~~
        b_j=\sum_{\alpha=0}^{p-1}\xi_j^\alpha  \;,
\label{q4}
        \ee
with $a_i|0\kt=b_j|0\kt=\zeta^\alpha_i|0\kt=\xi_j^\alpha|0\kt=0$ and
$i=1,\cdots,n_a$ and $j=1,\cdots,n_b$, for some positive integers
$n_a$ and $n_b$. The following relations express the relative
parastatistics of species of particles $a$ to $b$:
        \bea
        [\zeta_i^\alpha,\xi_j^\alpha]_{-\eta}&=&
        [\zeta_i^\alpha,\xi_j^{\alpha\dagger}]_{-\eta}\:=\:0\;,
\label{q5.1}\\
        \ll \zeta_{i}^{\alpha},\xi_{j}^{\beta} \rr_\eta&=&
        \ll \zeta_i^\alpha,\xi_j^{\beta\dagger}\rr_{\eta}\:=\:0\;,~~~(\alpha
        \neq\beta)\;,
\label{q5.2}
        \eea
where $\eta=\pm$ determines the relative statistics. $\eta=+$
(resp.~$\eta=-$) corresponds to the relative parabosonic
(resp.~ parafermionic) statistics.

The relative parastatistics is not determined by physical
reasoning, however there is a so called {\em normal relative
parastatistics} \cite{g-m} that generalizes the known case of $p=1$.
It is described as follows:
        \begin{itemize}
\item[I)] If $p_a\neq p_b$, then $a_i^\dagger,~a_i$
and $b_j^\dagger,~b_j$ anticommute if both particles $a$ and $b$ are
parafermions. Otherwise, they commute.
\item[II)] If $p_a=p_b=p$, then in Eqs.~(\ref{q5.1}) and (\ref{q5.2})
$\eta=-$, if both $a$ and $b$ are parafermions. Otherwise, $\eta=+$.
        \end{itemize}
The latter case says that two species of parafermionic particles
of the same order have parafermionic relative statistics. Whereas
a parabosonic particle has relative parabosonic statistics with respect
to both parabosonic and parafermionic particles of the same order.
This is a direct generalization of the $p=1$ case.

We conclude this section with a comment on the classical counterparts
of the quantum operators encountered above.

In the spirit of the work of Berezin \cite{berezin}, one defines
the classical analogs of $a_i^\dagger,~a_i$ and $\zeta_i^{\alpha\dagger},
{}~\zeta_i^\alpha$ as generators of algebras
defined by the rules given by (\ref{q1}) and (\ref{q2})
with the right hand side set to zero. Again the formula (\ref{q3})
establishes the relation between these algebras. The generators of
the former algebra, i.e., the one defined by setting the right hand side
of (\ref{q1}) to zero, with the sign ($-$) chosen in
Eqs.~(\ref{q1}), are called
{\em para-Grassmann variables of order $p$}. There is an alternative
definition of para-Grassmann variables advocated by Fillipov et al.
\cite{fillip} which is relevant to fractional supersymmetry. The latter
will not be employed in this article.

\section{The ($p=2$) Case and the Parabracket}
For $p=2$, the defining relations (\ref{q1}) simplify considerably
\cite{green,g-m}. One has:
        \bea
        a_ka_l^\dagg a_m\pm a_ma_l^\dagg a_k &=&
        2\delta_{kl}a_m\pm 2\delta_{lm}a_k\;,\nn\\
        a_ka_l a_m^\dagg\pm a_m^\dagg a_l a_k &=&
        2\delta_{lm}a_k\;,
    \label{q10}\\
        a_ka_la_m\pm a_ma_la_k&=&0\;,\nn
        \eea
which can be easily checked using the Green representation (\ref{q3}).
In these equations, the signs ($+$) and ($-$) correspond to ($p=2$)
parafermionic and parabosonic operators, respectively. Since we would like to
treat both of these operators simultaneously, the introduction of a
grading index $\mu=0,1$ is convenient, i.e., we attach $\mu$ to
operators $a_k$ and their Green components $\zeta_k^{\alpha}$ as a
superindex, and interpret $a_k^\mu$ and $\zeta_k^{\alpha\mu}$ as
parabosonic if $\mu=0$ and parafermionic if $\mu=1$. Now, we can
use $\mu$ to express the ($\pm$) signs in Eqs.~(\ref{q10}). In terms
of the Green components:
        \be
        a_i^\mu=\sum_{\alpha=0}^1 \zeta_i^{\alpha\mu}=
        \zeta_i^{0\mu}+\zeta_i^{1\mu}\;,
     \label{q11}
        \ee
one has:
        \bea
        [\zeta_i^{\alpha\mu},\zeta_j^{\alpha\mu\dagg}]_{(-1)^{\mu+1}}&=&
        \delta_{ij}\;,\nn\\
        {[}\zeta_i^{\alpha\mu},\zeta_j^{\alpha\mu}{]_{(-1)^{\mu+1}}}&=& 0\;,
    \label{q12}\\
        {[}\zeta_i^{\alpha\mu},\zeta_j^{\beta\mu\dagg}{]_{(-1)^\mu}}&=&
        {[}\zeta_i^{\alpha\mu},\zeta_j^{\beta\mu}{]_{(-1)^\mu}}\:=\: 0\,,~~~
        (\alpha\neq\beta)\;.\nn
        \eea
Note that throughtout the rest of this article the Green indices, $\alpha,
\beta,\cdots$, take values 0 and 1, for $p=2$.

It turns out that it is easier to work with self-adjoint (``real") operators
(variables). Thus we introduce yet another index $m=1,2$ and consider the
self-adjoint operators:
        \be
        \theta^{\alpha\mu}_{i1}:=\sqrt{\frac{\hbar}{2}}(\zeta_i^{\alpha\mu}
        +\zeta_i^{\alpha\mu\dagger})\;,~~
        \theta^{\alpha\mu}_{i2}:=-i\sqrt{\frac{\hbar}{2}}(\zeta_i^{\alpha\mu}
        -\zeta_i^{\alpha\mu\dagger}).
    \label{q15}
        \ee
Now, if one defines the {\em parabracket} by:
        \be
        \pbra
        \theta^{\alpha\mu}_{im},\theta_{jn}^{\beta\nu}
        \pket:=
        \theta^{\alpha\mu}_{im} \theta^{\beta\nu}_{jn}-(-1)^{
        \mu\nu+\alpha+\beta}
        \theta^{\beta\nu}_{jn}\theta^{\alpha\mu}_{im}\;,
      \label{q16}
        \ee
then the relation:
        \be
        \pbra
        \theta^{\alpha\mu}_{im},\theta_{jn}^{\beta\nu}
        \pket
        =\hbar\delta_{ij}\delta^{\alpha\beta}\ll
        i(1-\mu)(1-\nu)\epsilon_{mn}+\mu\nu\delta_{mn}\rr\;,
   \label{q16a}
        \ee
not only summarizes the defining relations (\ref{q2}) and hence (\ref{q1}),
but it also includes the statement of the normal relative parastatistics.
In Eq.~(\ref{q16a}), $\delta$ and $\epsilon$ are the Kronecker delta function
and the Levi Civita symbol, respectively.

One might view Eq.~(\ref{q16a}) as the statement of canonical quantization
for the ($p=2$) para-classical systems. In fact, the factor $\hbar$ has
been introduced so that (\ref{q16a}) yields the definition of the
classical counterparts of the quantum operators, i.e., ($p=2$)
parafermionic and parabosonic variables, in the limit $\hbar\to 0$.

The definition of the parabracket (\ref{q16}) may be extended to
polynomials in $\theta_{im}^{\alpha\mu}$. This is done by
defining it for the monomials, e.g.
        \be
        M:=\theta_{i_1m_1}^{\alpha_1\mu_1}\cdots
        \theta_{i_rm_r}^{\alpha_r\mu_r}\;,~~
        N:=\theta_{j_1n_1}^{\beta_1\nu_1}\cdots
        \theta_{j_sn_s}^{\beta_s\nu_s}\;,
   \label{mono}
        \ee
by
        \bea
        \pbra
        M,N
        \pket
        :=MN-(-1)^{\eta(M,N)}NM\;,
   \label{q20}\\
        \eta(M,N):=(\sum_{k=1}^r \mu_k)(\sum_{l=1}^s\nu_l)+
                r\sum_{l=1}^s\beta_l+s\sum_{k=1}^r\alpha_k\;,
   \label{q20b}
        \eea
and requiring bilinearity. In the classical limit, for any two polynomials
$P$ and $Q$ in $\theta_{im}^{\alpha\mu}$, one has
        \be
        \pbra
        P, Q
        \pket
        =0\;.
   \label{q20a}
        \ee

A more substantial result is a generalization of the Jacobi identity.
The following lemma can be easily proved by the application of
Eqs.~(\ref{q20}) and (\ref{q20b}).
        \begin{itemize}
        \item[ ]{\bf Lemma 1:} {\em
        Let $M,~N$, and $O$ be monomials in $\theta_{im}^{\alpha\mu}$
        and the parabracket $\pbra~,~\pket$, is defined by Eq.~(\ref{q20}),
        then the relation:
        \bea
        (-1)^{\eta(M,O)}\pbra M,\pbra N,O\pket\,\pket&+&
        (-1)^{\eta(O,N)}\pbra O,\pbra M,N\pket\,\pket\: +~~~
     \label{q31}\\
        &&
        (-1)^{\eta(N,M)}\pbra N,\pbra O,M\pket\,\pket\:=\: 0\;,\nn
        \eea
        holds as an identity.}\footnote{Eq.~(\ref{q31}) is a generalization
        of the the super-Jacobi identity \cite{bd} encountered in the
        study of supersymmetry. Thus it might be called the
        {\em para-Jacobi identity}.}
        \end{itemize}

Before proceeding further, we would like to make a further remark about
Eq.~(\ref{q16a}). This equation also provides a description of
the ($p=1$) case. This is done by making the Green indices vanish,
i.e., $\alpha,\beta,\cdots=0$.
This reveals the well-known fact that for the bosonic case ($\mu=\nu=0$),
the variables $\theta^{00}_{i2}$ correspond to the momenta
conjugate to the coordinates $\theta_{i1}^{00}$. This suggests
a similar pattern for the ($p=2$) case. That is, the
parabosonic coordinate variables in the Lagrangian formulation are
$\theta_{i1}^{\alpha 0}$. Whereas there is no such restriction on the
parafermionic variables. To demonstrate this in a unified notation, one
may introduce a collective index $I=(i;m)$, i.e., consider
$\theta_I^{\alpha\mu}$ , and require that for $\mu=0$,
$I=(i=1,\cdots,n_{\pi b};m=1)$ and for $\mu=1$, $I=(i=1,\cdots,n_{\pi f};
m=1,2)$, where $n_{\pi b}$ and $2n_{\pi f}$ are the number of
parabosonic and parafermionic degrees of freedom, respectively. In view of
this notation, one rewrites (\ref{q16a}) in the classical limit, as follows:
        \be
        \pbra
        \theta^{\alpha\mu}_{I},\theta_{J}^{\beta\nu}
        \pket
        =0\;.
   \label{q16b}
        \ee

One also must emphasize that the physical dynamical coordinate varaibles
are:
        \be
        \psi^\mu_I:=\sum_{\alpha=0}^1 \theta_I^{\alpha\mu}\;,
   \label{q17}
        \ee
and not the Green components $\theta_I^{\alpha\mu}$ themselves. In other
words, it is the algebra (ring) of polynomials ${\cal P}$ in $\psi^\mu_I$
that serves as the space of physical quantities. In view of Eq.~(\ref{q17}),
${\cal P}$ is a subalgebra (subring) of the algebra (ring) of
polynomials ${\cal T}$ in $\theta_I^{\alpha\mu}$. ${\cal T}$
is used as a larger space in which the calculations are performed.
To extract the physical results, one is bound to
project to the subspace ${\cal P}$. ${\cal P}$ and ${\cal T}$ have some
important subspaces. These are the even subalgebras ${\cal P}_2$ and
${\cal T}_2$, and the subalgebras generated by only the parabosonic
(parafermionic) variables $\psi^1_I$ (resp.~$\psi^0_I$) of ${\cal P}$,
and $\theta_I^{\alpha 1}$ (resp.~$\theta_I^{\alpha 0}$) of ${\cal T}$.
These are denoted by ${\cal P}^\mu$ and ${\cal T}^\mu$, respectively.

In view of Eq.~(\ref{q20a}), one finds that for example
the monomials in ${\cal T}_2$ either commute or anticommute.
In fact, if there is an even number of parafermionic factors
in an even monomial it commutes with all the even monomials and
two even monomials with odd numbers of parafermionic factors anticommute.
Furthermore, the even subalgebras of both ${\cal T}^\mu$
and hence ${\cal P}^\mu$, ($\mu=0,1$), are commutative. This is
important, because one would ordinarily like to choose ``physical"
quantities such as a Lagrangian to be a commutative object.
This cannot be fully achieved with ($p=2$) variables in general.
However, one might suffice to require that the Lagrangian be
chosen as a linear sum of even monomials each consisiting of an
even number of parafermionic factors. We shall offer a justification
for the latter requirement in Sec.~4.

In order to carry out the program of Lagrangian mechanics, one
also needs a differential calculus for the variables $\psi$'s or
alternatively for $\theta$'s. The latter also is addressed in the
earlier work in parastatistics \cite{o-k}. The results can be best
demonstrated using an extension of the definition of parabracket which also
applies
to ``partial derivatives":
        \bea
        \pbra
        \frac{\rdel}{\partial\theta_I^{\alpha\mu}},
        \frac{\rdel}{\partial\theta_J^{\beta\nu}}
        \pket
        &=&
        \pbra
        \frac{\ldel}{\partial\theta_I^{\alpha\mu}},
        \frac{\ldel}{\partial\theta_J^{\beta\nu}}
        \pket
        \:=\:0,
   \label{q21} \\
        \pbra
        \theta_I^{\alpha\mu},
        \frac{\rdel}{\partial\theta_J^{\beta\nu}}
        \pket
        &=&
        \pbra
        \frac{\ldel}{\partial\theta_I^{\alpha\mu}},
        \theta_J^{\beta\nu}
        \pket
        \:=\:
        \delta_{\mu\nu}\delta_{\alpha\beta}\delta_{IJ}\,,
   \label{q21a}
        \eea
where one defines the left hand sides of the latter equations
by replacing $\theta$'s in Eq.~(\ref{q16}) by either of
$\ldel\!\!/\partial\theta$ or $\rdel\!\!/\partial\theta$, with the same
indices.

Eqs.~(\ref{q21a}) may be used to obtain a generalized Leibniz rule.
One has:
        \begin{itemize}
        \item[ ] {\bf Lemma 2}: {\em
        Let $M$ and $N$ be monomials in ${\cal T}$ as given by
        Eq.~(\ref{mono}), then
        \be
        \frac{\rdel}{\partial\theta}(MN)=
        (\frac{\rdel}{\partial\theta}M)N-(-1)^{\eta(M,N)}
        (\frac{\rdel}{\partial\theta}N)M\;,
   \label{q30}
        \ee
        where $\eta(M,N)$ is defined by Eq.~(\ref{q20b}), and
        the indices of $\theta$'s are suppressed for simplicity.}
        \end{itemize}
A proof of Lemma 2 involves a lengthy two step induction on the
orders $r$ and $s$ of the monomials. Here, one makes extensive use
of Eqs.~(\ref{q16}), (\ref{q16b}) and (\ref{q21a}). Eq.~(\ref{q30}) is
of great practical use in performing computations with $\theta$'s.
A similar result may be proven for $\ldel\!\!/\partial\theta$.

We conclude this section with a discussion of the {\em reality}
condition. As is the case in the analysis of supernumbers \cite{bd},
we define a real element of the (complex) algebra ${\cal T}$, and
similarly ${\cal P}$, by introducing a $*$-operation. This is already
implicit in the quantum level in the definition of the Hermitian
conjugation. Following the ($p=1$) case \cite{bd}, we require
        \be
        \left( \lambda~
        \theta^{\alpha_1\mu_1}_{I_1}\cdots\theta^{\alpha_r\mu_r}_{I_r}
        \right)^*     :=
        \lambda^*
        \theta^{\alpha_r\mu_r}_{I_r}\cdots\theta^{\alpha_1\mu_1}_{I_1}
        \;,\nn
        \ee
and (additive) linearity of $*$-operation. Here $\lambda$ is a complex number
and $\lambda^*$ stands for its complex conjugate. A real element of ${\cal T}$
(resp.\ ${\cal P}$) is one whose $*$-conjugate equals itself.

\section{Peierls Bracket Quantization}
A generalization of Peierls bracket quantization to systems involving
bosonic (commuting) and fermionic (anticommuting) dynamical variables is
carried out in Ref.~\cite{bd}. Here a brief review is presented.

Consider a (non-relativistic) classical system whose dynamics is described by
the action functional:
        \be
        S[\Phi]:=\int_{\Phi} L(\Phi^i(t),\dot{\Phi}^i(t),t) dt \;,
     \label{q41}
        \ee
where $\Phi^i$ are the coordinate variables, $\dot{\Phi}^i$ are their
corresponding velocities, $t\in [0,T]$ is the time variable,
and $\Phi=(\Phi(t))$ is a path in the configuration space. Following
Ref.~\cite{bd}, let us denote the right and left functional derivatives
by
        \be
        S_{,i'}\equiv
        S[\Phi]\frac{\stackrel{\leftarrow}{\delta}}{\delta\Phi^i(t')}\;,
        ~~~_{i',}\!S\equiv
        \frac{\stackrel{\rightarrow}{\delta}}{\delta\Phi^i(t')} S[\Phi]\;,
     \label{q42}
        \ee
respectively. In this (condensed) notation, the indices represent both
the discrete and continuous (time) labels and repeated indices imply
summation over the discrete and integration over the continuous labels.
In particular, note that the prime on the index $i$ in (\ref{q42})
is associated with the  continuous index, $t'$.

The dynamical equations are given by
        \be
        S_{,i}=0\;.
      \label{q43}
        \ee
The second functional derivatives of the action functional yield the Jacobi
operator: $(_{i,}\!S_{,j'})$. The Green's functions of the latter
are defined according to their boundary conditions and the familiar relation:
        \be
        _{i,}S_{,j'}\,G^{j'k''}=-\,_{i}\delta^{k''}\;,
      \label{q44}
        \ee
where the repeated index $j'$ is summed and integrated over, and
        \[
        _{i}\delta^{k''} \equiv \delta_{i}^{k}\delta(t-t'')\;.
        \]

Denoting the advanced and retarded Green's functions by $G^+$ and $G^-$,
one defines the Peierls bracket of the fields $A=A[\Phi^i]$ and
$B=B[\Phi^i]$ according to:
        \be
        (A,B) := A_{,i}\, \tilde{G}^{ij'}\, _{j',}\! B\;,
     \label{q45}
        \ee
where the Green's function $\tilde{G}$ is defined by
        \be
        \tilde{G}:=G^+-G^- \;,
     \label{q46}
        \ee
It is called the {\em supercommutator function} by De~Witt \cite{bd}. One also
has the useful relation:
        \be
        (\Phi^i,\Phi^{j'}):=\tilde{G}^{ij'}\;.
     \label{q45a}
        \ee

The Peierls quantization scheme invloves the promotion of the classical
fields to linear operators acting on a Hilbert space and satisfying the
following (not necessarily equal time) supercommutation relations:
        \be
        [\hat{A},\hat{B}]_{\rm super}=i\hbar\widehat{(A,B)}\;.
     \label{q47}
        \ee
Here the hats are placed to emphasize that the corresponding quantities are
operators. They will be dropped where possible. If $A$ and $B$ have definite
parity, then the {\em superbracket} $[~,~]_{\rm super}$ becomes the ordinary
commutator if either of $A$ or $B$ is bosonic. Otherwise it becomes the
anticommutator. In practice, one usually uses Eq.~(\ref{q47}) written for
the coordinate variables, i.e.,
        \be
        [\hat{\Phi}^i,\hat{\Phi}^{j'}]_{\rm super}=
        i\hbar \widehat{\tilde{G}^{ij'}}\;,
        \ee
and properties of the Peierls bracket \cite{bd} (more conveniently those of
the superbracket) to compute the superbracket of other fields.

Employing the parity indices $\mu,~\nu,~\cdots$ of Sec.~2, i.e., considering
$\Phi^{i\mu}$, with $\mu=0$ corresponding to bosonic coordinates and
$\mu=1$ to the fermionic coordinates, one has
        \be
        [\hat{\Phi}^{i,\mu},\hat{\Phi}^{j'\nu}]_{\rm super}:=
        \hat{\Phi}^{i\mu}\hat{\Phi}^{j'\nu}-(-1)^{\mu\nu}
        \hat{\Phi}^{j'\nu}\hat{\Phi}^{i\mu}  \;.
      \label{q48}
        \ee
Ref.~\cite{bd} uses the same indices to label the coordinates and their
parity.

For this procedure to make sense, the Peierls bracket must possess a
series of properties. These are essentially the properties of
the supercommutator, namely the {\em supersymmetry} property:
        \be
         (A^\mu,B^\nu)=-(-1)^{\mu\nu}(B^\nu A^\mu) \;,
     \label{q49}
        \ee
and {\em super-Jacobi identity}:
        \be
         (-1)^{\mu\pi}(A^\mu,(B^\nu,C^\pi))
        +(-1)^{\pi\nu}(C^\pi,(A^\mu,B^\nu))
        +(-1)^{\nu\mu}(B^\nu,(C^\pi,A^\mu))  =0    \;,
     \label{q50}
        \ee
where $A^\mu,~B^\nu$ and $C^\pi$ are functions of $\Phi^i$ and have definite
parities $\mu,~\nu$ and $\pi$, respectively. Note that relations (\ref{q49})
and (\ref{q50}) are quite nontrivial. A proof of Eqs.~(\ref{q49}) and
(\ref{q50}) uses the symmetries of the Jacobi operator $_{i,}S_{,j'}$,
the supercommutator function $\tilde{G}^{ij'}$, and their functional
derivatives under the exchange of their indices \cite{bd}.

In view of the developments presented in the last section, we proceed
to generalize the Peierls scheme to systems involving ($p=2$)
parabosonic and parafermionic variables.\footnote{Inclusion of ordinary
fermionic and bosonic variables to such systems can also be carried out
within the framework presented in the present article.} In order to pursue
in this direction, we consider a Lagrangian $L$ built up of parabosonic
and parafermionic variables $\psi_I^\mu$ ($\mu=0,1$) and the corresponding
velocities, $\dot{\psi}_I^\mu$, i.e.,
        \be
        L=L(\psi^\mu_I,\dot{\psi}^\mu_I,t)\;.
        \label{q51}
        \ee
Note that the velocities are considered as independent
variables with the same parastatistical properties. In general, we shall
consider real Lagrangians which are even polynomials in both parafermionic
and parabosonic variables. The latter condition will prove essential
in having a consistent quantization scheme. For practical purposes, we
then switch to the Green's components $\theta_I^{\alpha\mu}$ and
$\dot{\theta}_I^{\alpha\mu}$. Using the calculus developed for
Green's components, one can define the notion of functional differentiation,
e.g., according to
        \bea
        F_{,i'}  &\equiv& F[\theta(t)]\frac{\stackrel{\leftarrow}{\delta}}{
                       \delta\theta^i(t')}\nn\\
                &:=& \lll
                F[\theta^1(t),\cdots,\theta^i
                (t+\epsilon\delta(t-t')),\cdots, \theta^d(t)]
                - F[\theta(t)] \rrr\left.
                \frac{\ldel}{\partial\epsilon}\right|_{\epsilon=0}\;,\nn
        \eea
where the index $i$ is a collective index representing $(I,~\alpha,
{}~\mu$) and $\epsilon$ is a variable with the same parastatistical
properties as $\theta^i$. The left functional derivative is defined
similarly.

Identifying the coordinate variables $\Phi^i$ of the beginning of
this section by $\theta^i$, with $i\equiv (I,\alpha,\mu)$,
the action functional, the dynamical equations, the Jacobi operator,
and its Green's functions are given according to Eqs.~(\ref{q41}),
(\ref{q43}) and (\ref{q44}), respectively. The {\em para-generalization}
of the Peierls bracket is obtained by Eqs.~(\ref{q45}) and (\ref{q46}). The
following analog of Eq.~(\ref{q47}) then yields the Peierls quantization
condition:
        \be
        \pbra \hat{A}, \hat{B}\pket= i\hbar \widehat{(A,B)}
     \label{q52}
        \ee
where $\pbra~,~\pket$ is the parabracket defined by Eq.~(\ref{q20}),
and $A$ and $B$ are polynomials in $\theta_I^{\alpha\mu}$. In particular,
one has:
        \be
        \pbra \hat{\theta}^i(t),\hat{\theta}^j(t')\pket =
        i\hbar \widehat{\tilde{G}^{ij'}}\;.
    \label{q53}
        \ee

The above procedure would be consistent provided that the para-generalized
Peierls bracket satisfies the symmetry properties of the parabracket,
namely the {\em parasupersymmetry} properties:
        \be
        (M,N)=-(-1)^{\eta(M,N)}(N,M)\;,
    \label{q54}
        \ee
and the {\em para-generalized Jacobi identity}:
        \be
        (-1)^{\eta(M,O)}(M,(N,O))+
        (-1)^{\eta(O,N)}(O,(M,N))+
        (-1)^{\eta(N,M)}(N,(O,M))=0\;,
    \label{q55}
        \ee
Here the function $\eta$ is the one defined by Eq.~(\ref{q20b}) and
$M,~N$, and $O$ are monomials in $\theta$'s.

A proof of Eq.~(\ref{q54}) follows from the following symmetry property
of {\em paracommutator function} $\tilde{G}^{ij'}$:
        \begin{itemize}
        \item[]{\bf Proposition~1}: {\em
        Let $\tilde{G}^{ij'}$ be defined by Eq.~(\ref{q46}),
        $i=(I,\alpha,\mu)$ and $j=(J,\beta,\nu)$, then:
                \be
                \tilde{G}^{ij'}=-(-1)^{\mu\nu+\alpha+\beta}
                \tilde{G}^{j'i}\;.
            \label{q56}
                \ee}
        \end{itemize}
To arrive at a proof of Prop.~1, we first state a couple of related results
which are labeled as Lemmas~3 and 4:
        \begin{itemize}
        \item[]{\bf Lemma~3}: {\em
        Let $M=\theta_{J_1}^{\gamma_{1}\rho_{1}}\cdots
        \theta_{J_D}^{\gamma_D\rho_D}$ be a monomial of order $D$,
        then:
                \be
                \frac{\rdel}{\partial\theta^{\alpha\mu}_I}M
                =(-1)^{\mu+\sum_{a=1}^{D}
                \eta(\theta^{\alpha\mu}_I,\theta^{\gamma_a\rho_a}_{J_a})}
                (M\frac{\ldel}{\partial\theta^{\alpha\mu}_I})\;.
            \label{q57}
                \ee}
        \end{itemize}
A proof of this result is obtained by a direct computation of both
sides of Eq.~(\ref{q57}) using the result of Lemma~2, i.e., Eq.~(\ref{q30}).
Next, we have:
        \begin{itemize}
        \item[]{\bf Lemma~4}: {\em
        Let M be as in Lemma~3, and let $i$ and $j$ label
        $(I,\alpha,\mu)$ and $(J,\beta,\mu)$ respectively,
        then
                \be
                \frac{\rdel}{\partial\theta^i}M\frac{\ldel}{\partial
                \theta^j}=
                (-1)^{(1+\sum_{a=1}^D\rho_a)(\mu+\nu)+\mu\nu+
                (D+1)(\alpha+\beta)}
                \frac{\rdel}{\partial\theta^j}M\frac{\ldel}{\partial
                \theta^i}\;.
            \label{q58}
                \ee
        In particular, if $M$ is an even monomial in both parabosonic
        and para\ -fermionic variables, then
                \be
                \frac{\rdel}{\partial\theta^i}M\frac{\ldel}{\partial
                \theta^j}=
                (-1)^{\mu\nu+\mu+\nu+\alpha+\beta}
                \frac{\rdel}{\partial\theta^j}M\frac{\ldel}{\partial
                \theta^i}\;.
             \label{q59}
                \ee}
        \end{itemize}
Lemma~4 is a straightforward consequence of Lemma~3. The statement of
Lemma~4 generalizes to the case of functional derivatives as well. Namely:
        \begin{itemize}
        \item[] {\bf Corollary}: {\em
        If the action functional $S$ consists of terms which are
        even monomials in both parabosonic and parafermionic variables,
        then one has:
                \be
                _{i,}\!S_{,j'}=(-1)^{\mu\nu+\mu+\nu+\alpha+
                \beta}~ _{j',}\!S_{,i}\;.
           \label{q60}
                \ee}
           \end{itemize}
Eqs.~(\ref{q44}) and (\ref{q60}) together with the observation that
both $_{i,}\!S_{,j'}$ and $G^\pm$ are even polynomials, lead to the
desired reciprocity relation:
        \be
        G^{\pm ij'}=(-1)^{\mu\nu+\alpha+\beta}G^{\mp j'i}
     \label{q61}
        \ee
This equation and the definition (\ref{q46}) yield a proof of Prop.~1.

Proof of the para-Jacobi identity (\ref{q55}) follows  essentially the
same procedure as in the ($p=1$) case \cite{bd}, but the computations are
more involved.

In the next section, we consider a simple example of application
of Peierls quantization program for a ($p=2$)--parafermionic system.

\section{One-dimensional Parafermi System}

Let us denote by $\psi$ a classical ($p=2$)--parafermionic (para-Grassmann)
variable with the Green components $\tau^\alpha:=\theta^{\alpha,\mu=1}_{I=1}$.
Then the defining relations (\ref{q10}), in the classical limit, imply
        \be
        \begin{array}{cc}
        \psi^3=\dot{\psi}^3=0\;,&\psi^2\dot{\psi}^2=\dot{\psi}^2\psi^2\;,\\
        \psi\dot{\psi}^2=-\dot{\psi}^2\psi\;,&
        \psi^2\dot{\psi}=-\dot{\psi}\psi^2\,
        \end{array}
     \label{q62}
        \ee
where $\psi$ and $\dot{\psi}$ are treated as independent ($p=2$)--parafermi
variables. Furthermore, one has the realtions:
        \be
        (\psi\dot{\psi})^2=(\dot{\psi}\psi)^2=0\;,
     \label{q63}
        \ee
which are most easily verified using the Green representation.

In view of Eqs.~(\ref{q62}) and (\ref{q63}), the most general real even
polynomial in dynamical variables -- upto an unimportant multiplicative
constant and additive total time derivatives -- has the form:
        \be
        L=\frac{A}{2}\psi^2+\frac{B}{2}\dot{\psi}^2+
          \frac{C}{4}\psi^2\dot{\psi}^2+ \frac{i}{4}(\psi\dot{\psi}-
          \dot{\psi}\psi)\;.
    \label{q64}
        \ee
Here $A,~B$ and $C$ are real numerical parameters. Eq.~(\ref{q64}) serves
as the most general possible form for the Lagrangian. In the following we
shall make a further demand, namely that the Peierls bracket quantization
and the canonical quantization of this system be consistent.

To carry out Peierls' program we first rewrite the Lagrangian (\ref{q64})
in terms of the Green components $\tau^\alpha$ and $\dot{\tau}^\alpha$
and compute the Jacobi operator. Here we suffice to state the results:
        \bea
        L&=&\sigma_{\alpha\beta}(
        \frac{A}{2}\tau^\alpha\tau^\beta+\frac{B}{2}\dot{\tau}^\alpha
        \dot{\tau}^\beta + \frac{C}{4}\sigma_{\gamma\delta}
        \tau^\alpha\tau^\beta\dot{\tau}^\gamma\dot{\tau}^\delta)
        +\frac{i}{4}\delta_{\alpha\beta}\tau^\alpha\dot{\tau}^\beta\;,
     \label{q65}\\
        _{\beta',}\!S_{,\alpha}&=&\left\{\left[\sigma_{\alpha\beta}
        (-B+\frac{C}{2}\sigma_{\gamma\delta}\tau^\gamma\tau^\delta)\right]
        \frac{\partial^2}{\partial t^2}+      \right.
        \label{q66}\\
                     && \left[ -i\delta_{\alpha\beta}
                               -C([(-1)^{\alpha+\beta}+1]\sigma_{\alpha
                               \gamma}\sigma_{\delta\beta}-
                               \sigma_{\alpha\beta}\sigma_{\gamma\delta})
                               \tau^\gamma\dot{\tau}^\delta\right]
        \frac{\partial}{\partial t}+\nn\\
                     && \left. \left[ A\sigma_{\alpha\beta}-C( [
                     \sigma_{\alpha\delta}\sigma_{\gamma\beta}
                     -\frac{1}{2}\sigma_{\alpha\beta}\sigma_{\gamma\delta}]
                     \dot{\tau}^\gamma\dot{\tau}^\delta+
                     \sigma_{\alpha\delta}\sigma_{\beta\gamma}
                     \tau^\gamma\ddot{\tau}^\delta)\right]
                     \right\}\delta(t-t')\;,\nn
        \eea
where $\sigma$ denotes the Pauli matrix $\sigma_1$, i.e.,
        \be
        \sigma_{\alpha\beta}=\left\{\begin{array}{cc}
                                1 & {\rm if}~~~ \alpha=\beta\\
                                0 & {\rm if}~~~ \alpha\neq \beta.
                                     \end{array}\right.
     \label{sigma}
        \ee

The Green's functions can be computed as power series in $(t-t')$,
similarly to the ($p=1$) case \cite{bd,p5}. A simple analysis of the
Green's functions, shows that if $B\neq 0$ or $C\neq 0$, then
$\pbra\tau^\alpha(t),\tau^\beta(t)\pket=0$, which is inconsistent with
the result of canonical quantization (\ref{q16a}), namely:
        \be
        \pbra\tau^\alpha(t),\tau^\beta(t)\pket= \hbar\delta^{\alpha\beta}\;.
     \label{q67}
        \ee
Setting $B=C=0$, and carrying out the computation of the Green's functions,
one finds:
        \be
        G^{\pm\alpha\beta'}=\left[
                \mp i\delta^{\alpha\beta}+ O(t-t')\right]\Theta[\pm(t-t')]
                \;,
        \ee
where $\Theta$ is the step function: $\Theta(t)=1$ if $t>0$,
$\Theta(t)=0$ if $t<0$, $\Theta(0)=1/2$.
The latter relation directly leads to Eq.~(\ref{q67}) and confirms the
consistency of the canonical and Peierls quantization programs.
Enforcing $B=C=0$ in the expression for the Lagriangian (\ref{q64}), one
has:
        \be
        L=\frac{i}{4}(\psi\dot{\psi}-\dot{\psi}\psi)+\frac{A}{2}\psi^2\;.
        \label{q68}
        \ee
The first couple of terms in the right hand side of (\ref{q68}) has the
same form as the kinetic term for ordinary fermionic systems. We shall
refer to these also as the kinetic term of the parafermionic system. The
last term serves as a potential term which does not have a counterpart
in fermionic systems.

A similar analysis shows that for ($p=2$)--parabosonic systems, choosing
the kinetic term to be of the same form as the bosonic kinetic term,
one ensures the consistency of the canonical and Peierls quantization
schemes.

\section{Conclusion}
The Lagrangian formulation of classical mechanics is shown to be applicable
to systems involving ($p=2$) parafermionic and parabosonic variables.
The introduction of the parabracket for the Green's components of the ($p=2$)
dynamical variables facilitates computations considerably. It also allows
for a generalization of the Peierls quantization program to such systems.

The internal consistency of the Peierls program requires the Lagrangian to
be an even polynomial in both parafermi and parabose variables.
The consistency of the results of the canonical and Peierls quantization
programs leads to the specification of the form of the parafermionic  and
parabosonic kinetic terms in the Lagrangian.

The material developed in this article has direct application in the
study of systems involving both the parafermi and parabose variables of order
($p=2$). Some examples of such systems have been encountered in the
context of parasupersymmetric quantum mechanics \cite{b-d}. There are still
quite a few unsettled issues regarding the true meaning of parafermi-parabose
(super)symmetry. Some of these issues are addressed in a companion paper
\cite{p11} using the formalism developed above.


\newpage

\begin{thebibliography}{99}
\bibitem{r-s} V.~A.~Rubakov and V.~P.~Spiridonov, Mod.~Phys.~Lett.~{\bf A3},
        1337 (1988).
\bibitem{b-d} J.~Beckers and N.~Debergh, Nucl.~Phys.~{\bf B340}, 767 (1990).
\bibitem{p8} A.~Mostafazadeh, ``Spectrum Degeneracy of General
        ($p=2$)--Parasupersymmetric Quantum Mechanics and Parasupersymmetric
        Topological Invarinats,'' Int.\ J.\ Mod.\ Phys.\ A, to appear (1995),
        hep-th/ 9410180.
\bibitem{frac} For example see: S.~Durand, Mod.~Phys.~Lett.~{\bf A8}, 2323
        (1993) and also \cite{fillip};
        For a more recent account: J.~A.~de Azc\'arraga and
        A.~J.~Macfarlane, ``Group Theoretical Foundations of Fractional
        Supersymmetry," Valencia Uni.~Preprint no: IFIC/95-23 (1995).
\bibitem{susy} L.~E.~Gendenshtein and I.~V.~Krive, Sov.~Phy.~Usp.~
        {\bf 28(8)}, 645 (1985); F.~Cooper, A.~Khare, and U.~Sukhatme,
        Phys.~Rep.~{\bf 251}, 267 (1995); and references therein.
\bibitem{witten} E.~Witten, Nucl.~Phys.~{\bf B202}, 253 (1982).
\bibitem{p9} A.~Mostafazadeh, ``Topological Aspects of Parasupersymmetry,''
        Submitted to Mod.~Phys.~Lett.~A (1995), hep-th/9506153.
\bibitem{khare} A.~Khare, J.~Phys.~A: Math and Gen. {\bf 25},  L749 (1992).
\bibitem{o-k} Y.~Ohnuki and S.~Kamefuchi, {\em Quantum Field Theory and
        Parastatistics,} Springer-Verlag, Berlin (1982).
\bibitem{green} H.~S.~Green, Phys.~Rev.~{\bf 90}, 270 (1953).
\bibitem{g-m} O.~W.~Greenberg and A.~M.~L.~Messiyah, Phys.~Rev.~{\bf 138},
        B1155 (1965).
\bibitem{a-m} F.~Ardalan and F.~Mansouri, Phys.~Rev.~{\bf D9}, 3341 (1974).
\bibitem{gro} A.~B.~Govorkov, Sov.~J.~Part.~Nucl.~{\bf 14}, 520 (1983).
\bibitem{goodman} M.~W.~Goodman, Commun.~Math.~Phys.~{\bf 107}, 391 (1986).
\bibitem{peierls} R.~E.~Peierls, Proc.~Roy.~Soc.~(London) {\bf A214}, 143
        (1952); See also \cite{bd}.
\bibitem{bd} B.~S.~De~Witt, {\em Supermanifolds}, Cambridge Uni.~Press,
        Cambridge (1992).
\bibitem{p5} A.~Mostafazadeh, J.~Math.~Phys.~{\bf 35}, 1095 (1994).
\bibitem{berezin} F.~A.~Berezin, {\em The Method of Second Quantization},
        Academic Press, New York (1966).
\bibitem{fillip} A.~T.~Filippov, A.~P.~Isaev, and A.~B.~Kurdikov,
        Int.~J.~Mod.~Phys.~{\bf A8}, 4973 (1993).
\bibitem{p11} A.~Mostafazadeh, ``Parabose-Parafermi Supersymmetry,''
        In preparation.
\end{thebibliography}
\end{document}